\documentclass{article}
\usepackage{times}
\usepackage{graphicx}

\title{An Exploratory Study of Personal Calendar Use}
\author{
  Manas Tungare, Manuel A. P\'erez-Qui\~nones, Alyssa Sams\\[4pt]
  Center for Human-Computer Interaction\\[2pt]
  Department of Computer Science\\[2pt]
  Virginia Tech\\[2pt]
  Blacksburg, VA, USA\\[2pt]
  \texttt{\{manas@tungare.name, perez@cs.vt.edu\}}
}
\date{}

\begin{document}
  \maketitle
\abstract{
  In this paper, we report on findings from an ethnographic study of how people use their calendars for personal information management (PIM). Our participants were faculty, staff and students who were not required to use or contribute to any specific calendaring solution, but chose to do so anyway. The study was conducted in three parts: first, an initial survey provided broad insights into how calendars were used; second, this was followed up with personal interviews of a few participants which were transcribed and content-analyzed; and third, examples of calendar artifacts were collected to inform our analysis. Findings from our study include the use of multiple reminder alarms, the reliance on paper calendars even among regular users of electronic calendars, and wide use of calendars for reporting and life-archival purposes. We conclude the paper with a discussion of what these imply for designers of interactive calendar systems and future work in PIM research.
}

\section{Introduction}
Personal Information Management (PIM) is receiving attention as an area of research within the CHI community \cite{barreau_2008_introduction, bergman_2004_personal, teevan_2006_personal}. PIM research mostly is concerned with studying how people find, keep, organize, and re-find (or reuse) information in and around their personal information space. Calendar management, one of the typical PIM tasks, is done today using a variety of systems and methods, including several popular paper-based methods: At-A-Glance, one of the largest suppliers of paper planners, sold more than 100 million calendars in 2000\footnote{http://www.allbusiness.com/consumer-products/office-supplies-equipment/6579063-1.html}.

For computer-based systems, calendar management is often integrated into email clients (e.g. Microsoft Outlook); it is one of the most common applications in all personal digital assistants (PDAs, e.g. Blackberries and iPhones), and there are several online calendar systems (e.g. Yahoo! Calendar, Google Calendar, Apple Mobile Me). Date- and time-based information is ubiquitous, and is often available through many means such as postings on office doors, displays with dated announcements, through email conversations, written on wall calendars, etc.  The result is that calendar information tends to be pervasive.

In this paper, we set out to explore how people use calendars in the presence of varied technological options. We are interested in understanding how calendar information is managed given the availability of these platforms. After a brief review of related work, we proceed to discuss our findings from the survey, interviews, and artifacts. From these, we suggest several opportunities for designers of future electronic calendar systems, and conclude the paper with a discussion of future research in personal information management.

\section{Related Work}

There is a long history of calendar studies in human-com\-pu\-ter interaction literature. Early research on calendar use predates electronic calendars. In 1982, Kelley and Chapanis \cite{kelley_1982_professional} interviewed 23 professionals to discover how people in the business world kept track of their schedules. They found that for the individuals interviewed, calendars were indispensable and showed a lot of diversity in their use. The use of multiple calendars was prevalent, and a wide variation was seen in the time spans viewed, as well as in other aspects such as archiving, editing and portable access. Many of the problems identified in paper calendars could be solved in electronic calendars, and they concluded with a list of features for emerging electronic calendars to implement. Soon afterwards, Kincaid and Pierre \cite{kincaid_1985_electronic} examined the use of paper and electronic calendars in two groups, and concluded that electronic calendars failed to provide several key features such as flexibility, power, and convenience, that paper calendars did. They recommended many useful features to be incorporated into electronic calendar systems as well.

Nearly 10 years after Kelley and Chapanis' original study, Payne \cite{payne_1993_understanding} conducted interviews with 30 knowledge workers about both calendars and to-do lists, followed by a task analysis of his observations. He concluded that the central task supported by calendars was \emph{prospective remembering}. Prospective remembering is the use of memory for remembering to do things in the future, as different from retrospective memory functions such as recalling past events. 

Payne reported that other uses of calendars were not as well-supported by systems at the time: features such as automatic scheduling were found to be detrimental, since it allowed users to schedule appointments without simultaneously being able to browse the other appointments in the calendar opportunistically. Electronic calendars of the time did not support the rich layout and typography of paper calendars, nor did they allow the scheduling of events that did not have a specific time associated with them. He concludes with a discussion how calendars could be designed to support the prospective remembering task better.

Groupware in the office or corporate context has been studied widely: Grudin \cite{grudin_1996_a-case} reported that many office administrators who used online calendar systems printed the calendars of their managers almost daily. Grudin and Palen \cite{grudin_1995_why-groupware, palen_1998_calendars, palen_2003_discretionary} explored the factors that contributed to adoption of groupware calendars, and the role of peer pressure and network effects in expanding the use of a shared calendar system for meeting scheduling. Palen \cite{palen_1999_social} explored the inter-play and co-evolution of individual demands, social aspects and deployment issues in groupware calendar systems. Calendars are often also used at home, but this area has not received much attention in the literature; a notable exception is \cite{crabtree_2003_informing}. In this paper, we focus on the personal use of calendars (at work or at home) but do not explore the collaborative aspects of calendar systems in detail.

Several interesting design concepts have been suggested to make electronic calendar systems less error-prone and smarter. These cover a wide range, from systems that retrieve tasks from email messages \cite{horvitz_1999_principles}, to systems that learn from users' behavior to recommend intelligent defaults \cite{mueller_2000_a-calendar}, to calendar systems that predict attendance at events \cite{tullio_2002_augmenting}. Beard et al. \cite{beard_1990_a-visual} assessed the effectiveness of a priority-based calendar prototype and concluded that integration with other personal information systems (such as email) would make the system more useful for users. Visualizing calendar information on desktop computers and mobile devices has been explored in several studies \cite{mackinlay_1994_developing, bederson_2004_datelens}.

In the field of Personal Information Management, the management of various information collections such as files \cite{barreau_1995_finding}, folders \cite{jones_2005_dont}, email \cite{whittaker_1996_email} bookmarks \cite{abrams_1998_information} and cross-collection issues \cite{boardman_2004_stuff, kaye_2006_to-have} have been studied widely. Calendars are an important part of users' personal information, and this domain can benefit from a re-examination in the wake of electronic and ubiquitous calendar systems. 

Jones \cite{jones_2004_finders} framed the problems in PIM in terms of the canonical tasks of \emph{keeping}, \emph{organizing}, and \emph{re-finding}. Keeping any kind of information involves a tradeoff between the likelihood it will be recalled in the future, and the costs of capturing and retaining it. Organizing involves filing it away such that it can be retrieved easily in the future, and while keeping the cost of organizing less than the cost of finding. Re-finding is a different problem from finding, since there are aspects to encountered information that make it personal.

With the increased use of mobile devices, more and more calendaring tasks are performed off the desktop computer. Perry et al. \cite{perry_2001_dealing} report on issues faced by mobile workers, their need for access to people and information located remotely, and the planful opportunism they engage in when utilizing their \emph{dead time} for tasks.

\section{Study Description}

The ethnographic approach we took in this study follows techniques commonly reported in the Personal Information Management literature, notably \cite{kelley_1982_professional, payne_1993_understanding, jones_2005_dont, marshall_2005_saving}. We did not attempt to test any \emph{a priori} hypotheses, but were interested in examining how calendar practices have evolved in the years following previous calendar studies by Kelley and Chapanis \cite{kelley_1982_professional} and Payne \cite{payne_1993_understanding}.

\begin{figure}[htbp]
  \centering
    \includegraphics[height=1in]{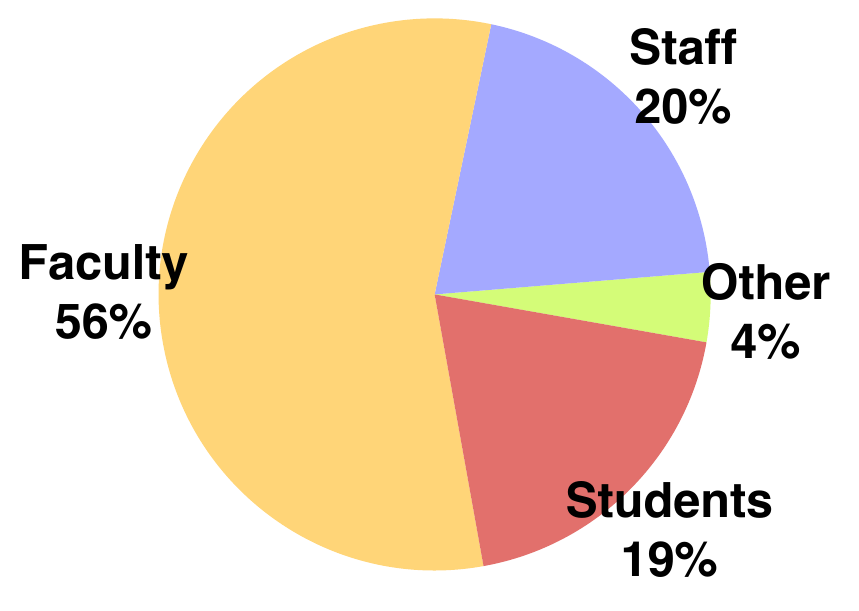}
  \caption{Roles of survey participants}
  \label{fig:roles}
\end{figure}

Our study has three components to it: a survey (N=98), in-person interviews (N=16), and an examination of calendar artifacts such as screenshots and paper calendars. A large-scale online survey was distributed among members of a university. A total of 98 responses were received (54\% male and 45\% female), including faculty (56\%), administrative staff (20\%), and students (19\%) (figure \ref{fig:roles}). While previous studies have examined organizational calendars \cite{dourish_1993_information} and groupware calendar systems \cite{grudin_1996_a-case, palen_2003_discretionary}, our focus was on the personal use of calendars.

In part two, we conducted in-depth personal interviews with 16 participants, recruited from among the survey participants. The recruitment criterion for interview candidates was the same as in \cite{kelley_1982_professional}: that participants should be regular users of some form of calendar system, either electronic or paper or a combination of both. Participants included graduate students, faculty members, administrative assistants, a department head, the director of a small business, etc., among others.

Interviews ranged from 20 to 30 minutes each, and were conducted \emph{in situ} at their workplaces so we could observe their calendaring practices directly (e.g. calendar programs or wall calendars or paper scraps.) Interviews were semi-structured and open-ended: a prepared set of questions was asked in each interview. The questions we asked were closely modeled on those asked in similar studies \cite{kelley_1982_professional, payne_1993_understanding}. The complete set of questions is available as an appendix in a technical report \cite{tungare_2008_an-exploratory}. As an extension to past studies, we were able to explore the use of features of modern calendar systems such as alarms, reminders, and mobile use, which were absent in paper calendars. Interviewees were encouraged to talk freely and to expand upon any of the themes they wished to discuss in more detail. Additional topics were addressed as appropriate depending on the interviewee's calendar use. Examining the calendar systems in use at their desks or on their walls prompted specific questions from the interviewers about these practices.

All interviews were transcribed in full. We performed content analysis \cite{krippendorff_2004_content} of the transcripts to extract common patterns of use. The main purpose of content analysis in this study was to summarize the findings into groups of common observations, as in \cite{marshall_2005_saving}. Individual responses were tagged into several categories by two of the authors and any differences reconciled by discussion. Nearly 410 tags resulted from this activity; these were then collapsed into 383 tags (grouping together tags that were near-duplicates) and 11 top-level groups during the clustering procedure. 

From each interview participant, we collected copies of artifacts that were used for calendaring purposes: 2 weeks' worth of calendar information and any other idiosyncratic observations that were spotted by the interviewers. These included screenshots of their calendar programs, paper calendars, printouts of electronic calendars (that were already printed for their own use), sticky notes stuck on paper calendars, etc. Some of these reflected a degree of wear and tear that occurred naturally over time; others provided evidence of manipulations such as color highlights, annotations in the margins, or comments made in other ways. Artifacts were not coded on any particular dimension, but pictures of these artifacts are used to supplant our textual descriptions wherever appropriate. (See figures \ref{fig:Church}, \ref{fig:StickyNoteOnCalendar} and \ref{fig:MoveCar}.)

\section{Analysis}
\subsection{General Use of Calendars}

\subsubsection{Capturing and Adding Events}
\label{sec:capture}
Capturing events refers to the act of knowing about an event and entering it into a calendaring system (also referred to as the `keeping' phase in the PIM literature.) Most survey participants reported adding new events as soon as they were (made) aware of them (93\%) while the rest added them before the end of the day. Even when at their desks, those users who owned PDAs reported using them to create new events in their calendar: this was deemed faster than trying to start the calendar program on a computer and then adding an event. When away from their desks, they used proxy artifacts such as printed calendar copies or paper scraps.

\begin{figure}[htbp]
  \centering
    \includegraphics[height=1.65in]{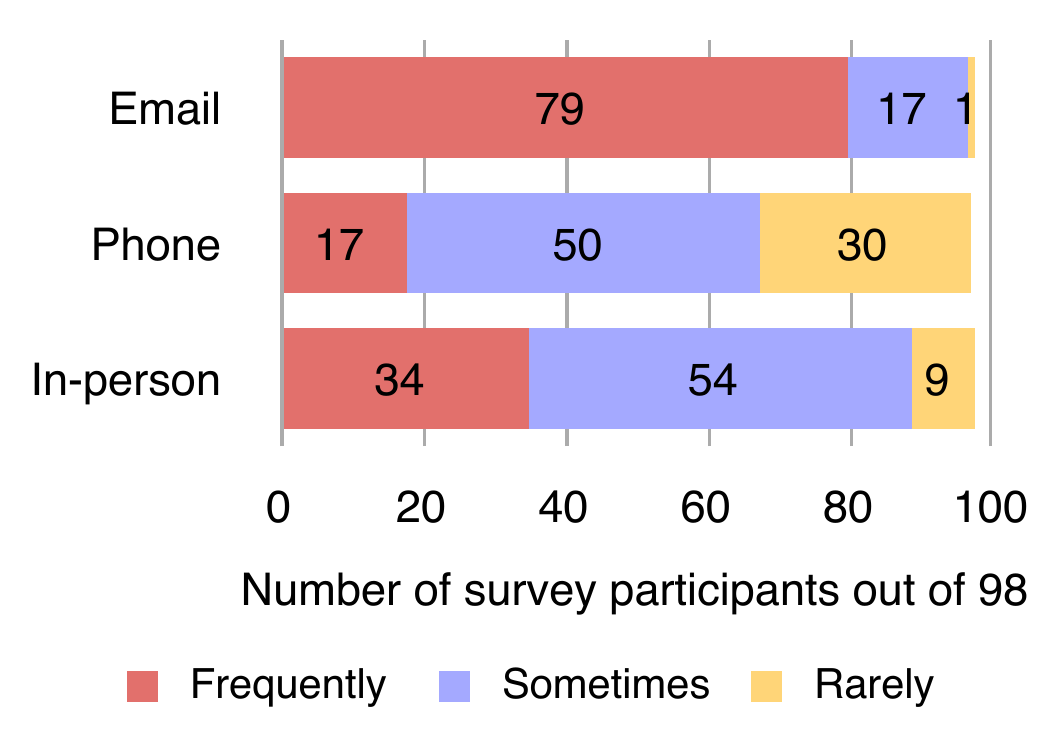}
  \caption{How new events arrive in users' PIM systems}
  \label{fig:NewEventsArrive}
\end{figure}

Information about new events reached the primary calendar user via one of several means: email, phone, and in-person were commonly reported (figure \ref{fig:NewEventsArrive}). The fact that email was the most common way reported in our study is an expected evolution from older findings \cite{kelley_1982_professional} that phones were the most common stimuli for calendar events. Interviewees mentioned several other methods through which they received events: flyers, posters, campus notices, meeting minutes, public calendars (such as academic schedules or sports events), newspapers, internet forums, (postal) mail, fax, radio, or scheduled directly by other people who had access to the calendar (e.g., shared calendars). The wide variety of sources here is a potential indication of the problem of information overload \cite{schick_1990_information} faced by knowledge workers.

\subsubsection{Personal Calendar View Preference}

\begin{figure}[htbp]
  \centering
    \includegraphics[height=1.35in]{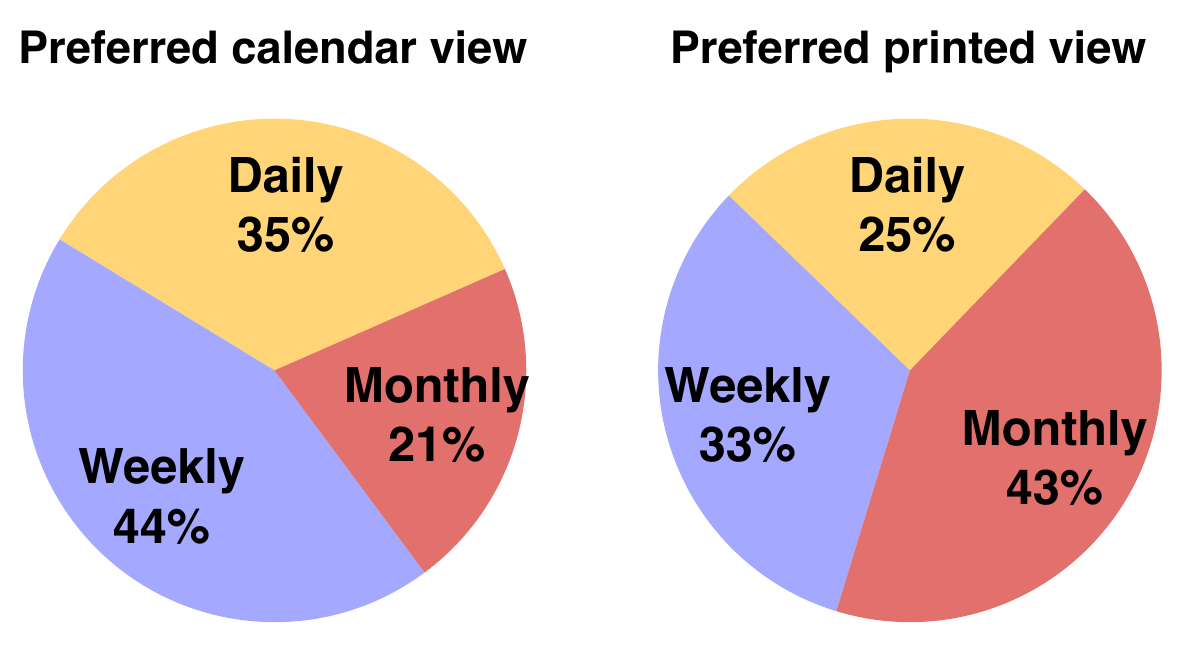}
  \caption{Preferred calendar views}
  \label{fig:preferred-views}
\end{figure}

We refer to the most common time interval shown in a calendar program or on a paper calendar as the \emph{preferred personal calendar view}: the week view was preferred by most of our survey participants at 44\%, followed by the day view at 35\%, and the month view at 21\% (figure \ref{fig:preferred-views}). These are very close to the numbers reported by Kelley et al. \cite{kelley_1982_professional} (45\%, 33\%, 22\% respectively).

That many interviewees preferred a week view suggests the use of the calendar for \emph{opportunistic rehearsal}, because they browsed the entire week's appointments each time they view\-ed the calendar. This preference supports the analysis of \cite{payne_1993_understanding} in that the printed versions of calendar do provide a valuable aid in opportunistic reading of the the week's activities. Users who kept multiple calendars within the same calendaring system indicated that they turned the visibility of each calendar on or off on demand, based on the specifics of what they needed to know during a particular lookup task. On smaller devices such as PDAs, the default view was the daily view.

There seem to be two motivators for browsing calendars: looking for activities to attend in the near future, and looking for activities further out that require preparation. A daily view directly supports the first, while a week view partially supports the second one. Intermediates such as Google Calendar's 4-day view afford browsing for future events without losing local context for the current day. The downside of such a view, however, is that days no longer appear in a fixed column position, but in different locations based on the day. Thus, the preferred calendar view depends on the type of activity the user is doing.

\subsubsection{Frequency of Consulting the Calendar}
When asked about the frequency at which users consulted their calendars, we received a wide range of responses in the survey: keeping the calendar program always open (66\%) and several times a day (21\%) were the most common.

In the interviews, several other specific times were reported: just before bedtime or after waking up; only when prompted by an alarm; when scheduling a new event; once weekly; or on weekends only. Two interviewees reported consulting their calendar only to check for conflicts before scheduling new events, and for confirmation of events already scheduled. 

\subsubsection{Proxy Calendar Artifacts}
We use the term \emph{`proxy calendar artifacts'} (or \emph{`proxies'} in short) to refer to ephemeral scraps or notes (characterized as micronotes in \cite{lin_2004_understanding}) or printed calendars or electronic means such as email to self that are used for calendaring when primary calendar systems are unavailable or inaccessible (e.g. when users were away from their desks or offices). 

Despite the prevalent use of electronic calendars, many were not portable and were tied to specific desktop computers. This prompted the users to use other means to view or add events to their calendar; about 27\% reported that they used proxy artifacts such as scraps or notes to be entered into the primary calendar at a later time. A wide variety of proxy calendar artifacts was reported in our interviews: paper scraps were by far the most common medium; other techniques included carrying laptops solely for the purpose of calendaring, PDAs, voice recorders, and printouts of electronic calendars. Information captured via these proxies was transferred to the primary calendar after a delay: most often, users entered the events as soon as they could access their primary calendar (63\% of survey participants), a few others reported entering them within the same day (25\%), while the maximum delay reported was up to one week.

\subsubsection{Information Stored in an Event Record}
Calendar systems allow users to add several items of information to an event record. Typical information included the date of the event (97\%), time (96\%), location (93\%) and purpose (69\%) as indicated in the survey. In interviews, it was clear that common fields such as notes, other attendees and status were used only to a limited extent. Location was entered mostly for non-recurring events. However, many other pieces of information were frequently recorded, even though calendar programs do not have a specific field for these data. For example, information critical for participation at an event was entered inline for easy access: e.g. phone numbers for conference calls, cooking menus and shopping lists, meeting agenda, original email for reference, links to relevant web sites, and filenames of relevant files. 

One participant mentioned adding meeting participants' email addresses in case she needed to inform them of a cancellation or rescheduling. For activities such as trips or flights, further details such as booking codes and flight details were included as a way of reducing information fragmentation between the calendar system and the email system.

\subsubsection{Types of Events}
The events most commonly recorded on calendars by survey participants were timed events such as appointments or meetings (98\%), special events requiring advance planning, such as tests (93\%), long duration events such as the week of final exams at the end of each semester (66\%), and all-day events such as birthdays (81\%). Several interviewees also mentioned recording to-do items in a calendar, such as phone calls to be made, or tasks which would remain on the calendar until completed, or which were scheduled in on their deadline. Specifically, we found several instances of the following types of events scheduled:

\begin{itemize}
  \item \textbf{Work-related events.} Many interviewees used calendar scheduling for work-related events such as meetings, deadlines, classes, public events such as talks and conferences, and work holidays. Users in work environments included vacation details for co-workers and subordinates. Time was routinely blocked off to prepare for other events: e.g. class preparation or ground work to be done before a meeting.

Interviewees who had administrative assistants reported that their assistant maintained or co-maintained their calendar (7 out of 16 interviewees). The dynamics of shared access were vastly different across all these situations. One interviewee mentioned that he would never let an assistant be their primary scheduler; the assistant was able to access only a paper copy and any new events would be reviewed and added by the primary calendar user. Two other users mentioned that they provided paper calendars to subordinates to keep track of their schedule and to be able to answer questions about it to third parties. One participant reported calling in to their secretary when they needed to consult their schedule while away from their desk (similar to previous reports in \cite{perry_2001_dealing}), while another reported sending email to themselves as a way to quickly capture a newly-scheduled meeting.

  \item \textbf{Family/personal events.} Half of the survey respondents indicated that they coordinate calendars with their spouses, roommates, or family. Even though family activities such as picking up kids from school, or attending church services, were easily remembered without the aid of a calendar, interviewees reported that they chose to record them anyway to provide \emph{``a visual idea of the entire day''} (figure \ref{fig:Church}). Public holidays, family birthdays, and guest visits were added to prevent accidental scheduling of conflicting events. 

  \begin{figure}[htbp]
    \centering
      \includegraphics[height=1.9in]{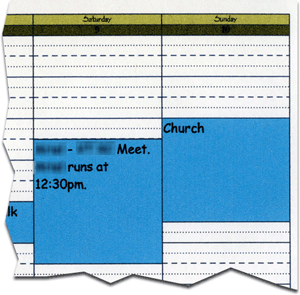}
    \caption{Family events such as attending church are added to calendars, not for remembering, but to be able to get a visual idea of the entire day.}
    \label{fig:Church}
  \end{figure}

  Many participants reported having separate calendars for business use and for home/personal use, as was also seen in a majority of respondents in \cite{kelley_1982_professional}. Although events overlapped between them (e.g. work trips on family calendars and family medical appointments on work calendars), the calendars themselves were located at the respective places and maintained separately. Family calendars were most likely to be kept in the kitchen, on the refrigerator.
  
  Two contrasts between work calendars and home calendars were prominent: work calendars were more often electronic while home calendars more likely to be paper calendars, e.g. as a wall calendar, or on the refrigerator. Work calendars were updated by the primary users or their secretaries or their colleagues, while family calendars were overwhelmingly managed by women. No male participant reported being the only calendar manager at home; women reported either being the only person to edit it, or sharing responsibilities with their husbands.

  Family-related events and reminders were constrained to the home calendar, as in \cite{nippert-eng_1996_calendars}, but they were sometimes added to work calendars if such events would impact work time. For example, medical appointments (of self or family members) that occurred during work hours were added to work calendars so that their co-workers were aware of their absence.

  \item \textbf{Public events.} Public events were added even when the user had no intention of attending that event. They were added to recommend to other people, or for personal planning purposes, or to start conversations related to the public activity. An administrator (from 
  ANONYMIZED,
  a small university town with a very popular college football team) said that although he had no interest in football, he added home games to his calendar to ensure that visiting dignitaries were not invited during a time when all hotels in town would be booked to capacity. On the other hand, two interviewees considered such public events as contributing to clutter in their personal calendar, and chose not to add them.

\end{itemize}

\subsection{Continued Use of Paper Calendars}

In his 1993 study \cite{payne_1993_understanding}, Payne reports that the most stable characteristic he observed was the continued reliance of all but two participants on some kind of paper calendar. Our findings are similar: despite most of our users using electronic calendars, every one of them reported using paper calendars even if not regularly; 12 out of 16 interview participants reported using them regularly.

\subsubsection{Reasons for the Continued Use of Paper Calendars}
We group the several reasons and examples elicited from our participants into the following four categories:

\begin{itemize}
  \item \textbf{Paper trail.} Cancelled events were scratched off the calendar, leaving a paper trail. Being able to make a distinction between cancelled and never-scheduled events was cited as an important concern for continuing with paper calendars.

  \item \textbf{Opportunistic rehearsal.} We found support for the idea of opportunistic rehearsal \cite{payne_1993_understanding}. Users cited that wall calendars needed no more than a glance to read, and provided for quick reference. This also corroborates Dourish's argument \cite{dourish_1993_information} that the presence of informational context in paper artifacts such as calendars is an important motivator for people to continue to use them, even though electronic systems support the information retrieval task better.

  \item \textbf{Annotation.} Paper calendars are more amenable to free-form annotation, as reported earlier \cite{kelley_1982_professional}, and as the following quotes from our study illustrate:
  \begin{quote}
    \emph{``That's what I call the graffiti aspect of it, it's probably freer by virtue of being handwritten.''}
    \\[8pt]
    \emph{``There is a lot of that [code and symbols]. Stars and dashes and circles and headlines, marked and completed.''}
  \end{quote}

  Figure \ref{fig:StickyNoteOnCalendar} shows a printed calendar with a sticky note pasted on it. The event is about a community potluck dinner. The sticky note complements the scheduled appointment with information about the dish the participant plans to bring to the event. Figure \ref{fig:Halloween} shows a picture of a pumpkin hand-drawn on a printed calendar to mark Halloween on October 31.

\begin{figure}[htbp]
  \centering
    \includegraphics[height=1.9in]{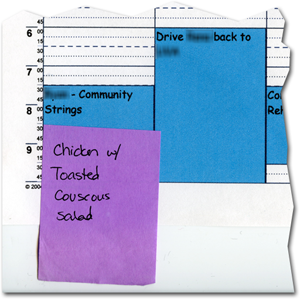}
  \caption{Sticky notes are pasted on paper calendars to remind oneself of the preparation required for an event.}
  \label{fig:StickyNoteOnCalendar}
\end{figure}

\begin{figure}[htbp]
  \centering
    \includegraphics[height=2in]{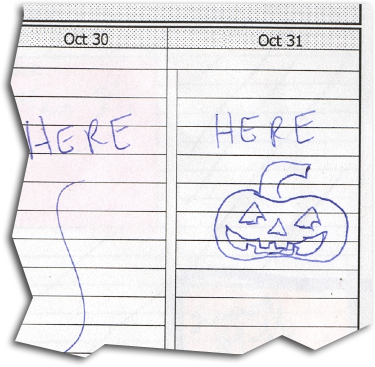}
  \caption{A hand-drawn pumpkin marks Halloween on October 31.}
  \label{fig:Halloween}
\end{figure}

  \item \textbf{Prepopulated events.} Participants reported that having holidays or other event details already printed in com\-mer\-cial\-ly-avai\-la\-ble  paper calendars was an important reason for using them. Calendars distributed by the university contained details not only of academic deadlines, but also of athletic events and games; \cite{kelley_1982_professional} point to branding issues as well.

\end{itemize}

Paper calendars were used alongside electronic calendars in either a supplementary or complementary role, as follows:

\subsubsection{Printouts of Electronic Calendars}

\begin{figure}[htbp]
  \centering
    \includegraphics[width=0.47\textwidth]{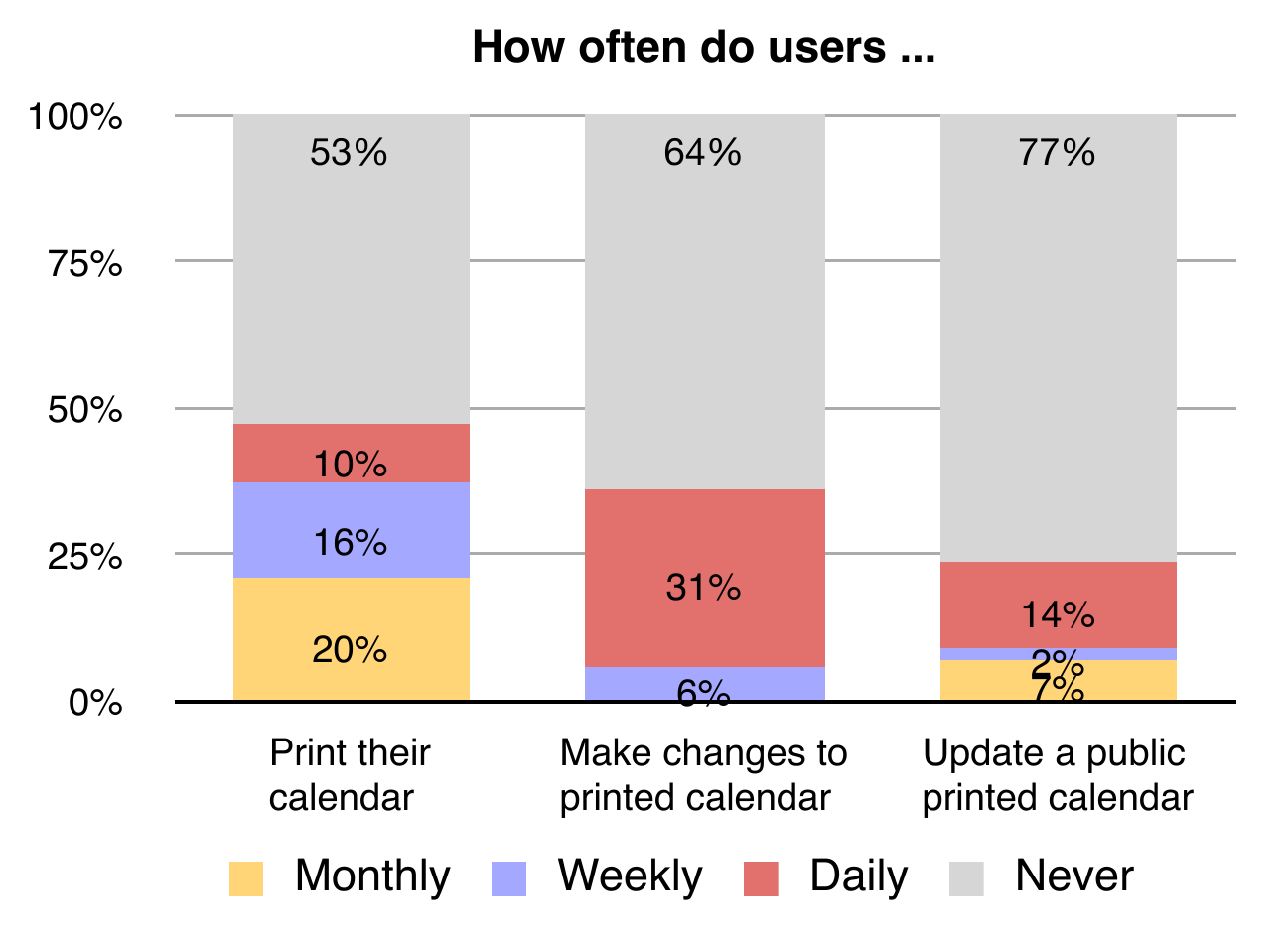}
  \caption{How often users perform activities related to paper calendars.}
  \label{fig:how-often-do-users}
\end{figure}

Printouts of electronic calendars played a supplementary role: they were used as proxies of the master calendar when the master calendar was unavailable. 35\% of survey participants reported printing their calendar. Among those printed, all views were commonly printed: monthly (43\%), weekly (33\%) and daily (25\%) (figure \ref{fig:preferred-views}). Among those who printed, many printed it monthly, weekly or daily (figure \ref{fig:how-often-do-users}).

Based on our interviews, we found that electronic calendars were printed for three main reasons:

\begin{itemize}
  \item \textbf{Portability.} Users carried a printed copy of the master calendar to venues where collaboration was anticipated, such as a meetings or trips. Even those who carried laptops and PDAs said that they relied on printed calendars for quick reference.
  
  \item \textbf{Quick capture.} Events were often entered into paper calendars first because of their easy accessibility, and were later transferred back to the digital calendar. \ref{sec:capture} One-third of all interviewees reported making changes to paper copies of their calendars. Not all these changes were propagated back to the master calendar, however.

  \item \textbf{Sharing a read-only view with associates.} Taping a prin\-ted calendar to the outside of office doors was common practice, as reported by interviewees. In one instance, a user provided printed calendars to his subordinates so they could schedule him for meetings. These events were then screened by him before being added to the master calendar.

\end{itemize}

\subsubsection{Wall Calendars}

Wall calendars typically played a complementary role, and there was little overlap between the events on a wall calendar and those in an electronic calendar. 70\% of survey participants had a wall calendar in their home or office, however only 25\% of users actually recorded events on it. Family events such as birthdays, vacations, and days off were most commonly recorded by interviewees. At home, wall calendars were located in the kitchen, on the fridge.

\subsubsection{Index Cards}

An extreme case of ad hoc paper calendar usage reported by one of our interviewees involved index cards, one for each day, that the participant carried in his shirt pocket when he forgot his PDA. Another interviewee reported exclusively using index cards for calendar management at their previous job because of their portability and trustworthiness.

\begin{quote}
  \emph{``[...] I get a 3$\times$5 index card so that I can stick it in my pocket without me losing it. Because if it is put on a little piece of paper, I am sure I will lose it. But a 3$\times$5 index card, it fits in my pocket, but is big enough so I would not lose it.''}
\end{quote}

\begin{quote}
  \emph{``I used to just take index cards and keep a little stack [...] clipped together and keep it in my back pocket. I got my schedule, I have my daily schedule and each index card was a day so I would write in the daily events [...]. I could easily carry over events from previous days because I just move that card. Everybody joked about it but it was a really useful system.''} (edited to remove conversational pauses and fillers.)
\end{quote}
 
\begin{quote}
  \emph{``One for the paper, again, I know if they are going to blow up or not. This [referring to index cards] I am still waiting for it to explode on me some day.''}
\end{quote} 

We report this not as a trend, but to illustrate the wide variety in the use of paper calendars.

\subsection{Reminders and Alarms}

Reminders and alarms are one of the major distinguishing features of modern electronic calendar systems. A majority of survey participants (63\%) reported using these features. One user reported switching from paper to an online calendar because \emph{``a paper calendar cannot have an alarm feature''}. We use the term \emph{reminder} to refer to any notification of a calendar event, and \emph{alarm} to refer to the specific case of an interruption generated by the calendar system. Based on our interviews, we classified reminders into three categories taking into consideration the reasons, time, number, modalities and intervals of alarms. Before presenting the details of such a classification, however, we examine the individual factors in more detail.

\subsubsection{Reasons for Using Alarms}

Although reminding oneself of upcoming events is the most obvious use case for alarms, there were several other situations where users mentioned using reminders in addition to consulting their calendars regularly. Even when users were cognizant of upcoming events, they preferred to set alarms to interrupt them and grab their attention at the appointed hour. Alarms served as preparation reminders for events that were not necessarily in the immediate future.

\begin{figure}[htbp]
  \centering
    \includegraphics[width=0.45\textwidth]{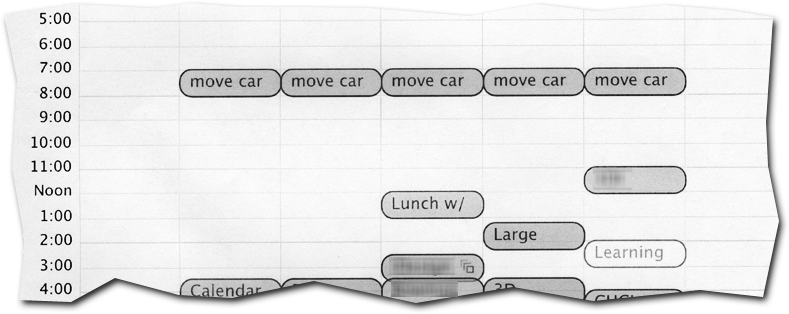}
  \caption{One user showed us an example where he set an alarm for moving his car before a towing deadline set in at 8:00am.}
  \label{fig:MoveCar}
\end{figure}

When subordinates added events to a primary user's calendar, alarms were deemed an important way of notifying that user of such events. Early morning meeting reminders doubled up as wake-up alarms: one interviewee reported keeping their PDA by their bedside for this purpose. Another interviewee who needed to move his car out of a university parking lot where towing started at 8:00 am sharp had set a recurring alarm (figure \ref{fig:MoveCar}). In one case, alarms were closely monitored by a user's secretary: if an event were missed by the user by a few minutes, the secretary would check on her boss and remind him to attend the meeting that was now overdue. 

\subsubsection{Number and Modalities of Reminders}
While most survey participants only set a single reminder per event (52\%), many others reported using multiple alarms. We conclude from our interviews that different se\-man\-tic mea\-nin\-gs were assigned to each such reminder: an alarm one day before an event was for preparation purposes, while an alarm 15 minutes before an event was a solicited interruption. Multimodal alarms were not used by many: the two most popular modalities used individually were audio (40\%) and on-screen dialogs (41\%).

\subsubsection{Alarm Intervals}
\label{sec:alarm-intervals}

Reminders were set for varying intervals of time before the actual events took place, ranging from 5 minutes to several years. The two factors that affected this timing were (1) location of the event, and (2) whether or not (and how much) preparation was required. Users often set multiple alarms to be able to satisfy each of these requirements, because a single alarm could not satisfy them all. Based on these findings, we classify alarms into 3 categories:

\begin{itemize}
  \item \textbf{Interruption Reminders.} Alarms set 5-15 minutes before an event were extremely short-term interruptions intended to get users up from their desks. Even if they knew in their mind that a particular event was coming up, it was likely that they were involved in their current activity deeply enough to overlook the event at the precise time it occurred. 15 minutes was the most common interval, as reported by 8 out of 16 interview participants.

  We found that the exact interval for interruption reminders was a function of the location of the event. Events that occurred in the same building as the user's current location had alarms set for between 5 and 15 minutes. Events in a different building had alarms for between 15 minutes and 30 minutes, based on the time it would take to reach there. Two interviewees reported that they set alarms for TV shows and other activities at home for up to 1 hour prior, because that is how long their commute took. 

  \item \textbf{Preparation Reminders.} Users set multiple alarms when preparation was required for an event: the first (or earlier) alarm was to alert them to begin the preparation, while a later alarm was the interruption reminder for that event. Payne \cite{payne_1993_understanding} mentions the prevalence of this tendency as well: the reason for the first alarm (out of several) is to aid prospective remembering where the intention to look up an event is not in response to a specific temporal condition, but instead such conditions are checked after the intention is recalled. If certain items were needed to be taken to such meetings, preparation reminders were set for the previous night or early morning on the day of the event. Based on the interviews, preparation reminders were more commonly used for non-recurring events than for recurring events.

  \item \textbf{Long-term Reminders.} Events several months or years into the future were assigned reminders so that the user would not have to remember to consult the calendar at that time, but instead would have them show up automatically at (or around) the proper time. This is an illustration of using the calendar for prospective remembering tasks. Examples include a department head who put details of faculty coming up for tenure in up to 5 years, and a professor setting reminders for a conference submission deadline several months later.

\end{itemize}

\subsection{Calendars as a Memory Aid}

Calendars serve a value purpose as external memory for ev\-ents \cite{payne_1993_understanding}.
In addition, in our data we found that the role that calendars play with respect to memory goes beyond this simple use. In particular, the following uses of calendars illustrate the different ways in which calendars serve as memory aids beyond simple lookups: First, users reported recording events in the calendar after the fact, not for the purpose of reminding, but to support reporting needs. Second, a few reported using previous years' calendars as a way to record memorable events to be remembered in future years. For those that used paper calendars, these events were often copied at the end of the year to newer calendars. The function of memory aid goes beyond remembering personal events (appointments and deadlines); it serves as a life journal, capturing events year after year. Kelley and Chapanis \cite{kelley_1982_professional} reported that 9 out of 11 respondents in their study kept calendars from two to 15 years.

\subsubsection{Reporting Purposes}
\label{sec:reporting-purposes}

In our study, 10 out of 16 interviewees reported that they used their calendar to generate annual reports every year. Since it contained an accurate log of all their activities that year, it was the closest to a complete record of all accomplishments for that year. Among these, 5 users reported that they archived their calendars year after year to serve as a reference for years later. This tendency has also been reported in past studies \cite{kelley_1982_professional, payne_1993_understanding}; Kelley referred to it as an `audit trail', and highlighted the role of calendars in reporting and planning.

One person mentioned that they discovered their father's journal a few years after his death, and now they cultivate their calendar as a memento to be shared with their kids in the future.

\begin{quote}
  \emph{``I think I occasionally even think about my kids. Because I do, I save them, I don't throw them away [...] I think that it's common with a little more sense of mortality or something. It's trying to moving things outwards.''}
\end{quote}

\section{Opportunities for Design}

In this section, we highlight how some of our findings can be address through new electronic calendar designs.

\subsection{Paper Calendars and Printing}

We do not believe that paper calendars will disappear from use; they serve several useful functions that are hard to replace by technology. Electronic calendars in general are more feature-rich than paper calendars. Portable devices have good support for capturing information while mobile. Yet, we found that paper calendars and proxies continue to be prevalent in the use of calendar management. They provide support for easy capture of calendar information, are effective at sharing, and support the display of the calendar in public view with ease.

Therefore, given the many uses of paper calendars, we consider how electronic calendar systems can provide better support for these proxies. Richer printing capabilities might provide easy support for transferring online calendar information to the paper domain. Printing a wall calendar is a novelty relegated to specialized design software. However our findings show that wall calendars play a significant role in supporting calendar management, particularly at home. With affordable printing technology available, it is possible to print a wall calendar or table calendar at home, incorporating not only details of events from a user's personal electronic calendar, but also visual elements such as color coding, digital photos (for birthdays, etc.) and event icons. In a way, printed calendars are used in similar ways as discussed in \cite{lin_2004_understanding}.

\subsection{Digital Paper Trails}
Some of the features of paper calendars can be recreated in online systems. For example, current electronic calendar systems remove all traces of an event upon cancellation, without providing an option to retain this historical record. This was one of the shortcomings which led interview participants to rely on paper instead. Instead of deleting events, they could be faded out of view, and made visible upon request.  Most calendar software support the notion of different calendars inside of the same program. A possibility is that all deleted events could simply be moved to a separate calendar, where events can be hidden easily. Yet, the events would remain in the calendar as a record of cancelled activity.

\subsection{Tentative Event Scheduling}
Several participants indicated that they `penciled in' appointments in their paper calendars as tentative appointments to be confirmed later (also identified as a problem in \cite{kelley_1982_professional}). These tentative appointments served as a way of blocking particular date/time combinations while a meeting was being scheduled with others. Often, there were several of these tentative times for a particular meeting. Once the meeting was confirmed, only one of them was kept and the rest discarded. This type of activity is not well-supported in personal calendars. For corporate calendars, there is adequate support for scheduling group meetings, but it is often missing in personal calendars.

\subsection{Intelligent Alarms}
Calendar alarms and reminders have evolved from past systems and now allow notification in several ways: audible alarms, short text messages, popup reminders, and email are just a few. However, the fundamental concept of an alarm still tailors only to interruption reminders.

\begin{itemize}
  \item \textbf{Preparation reminders.} To support pre\-pa\-ra\-tion re\-min\-ders, many electronic calendars allow the creation of multiple alarms per event, with different modalities for each (e.g., email, SMS, sounds, dialog box). However, when these reminders are used for preparation, as we found in the study, users often wanted to have more context: they expected to have an optional text note to indicate what preparation was required. E.g., alarms that would remind a user before leaving home to remember to carry material for an upcoming meeting, or a reminder the previous night to review documents.

  \item \textbf{Location-related alarms.} The location of events was found to be an important influencer of alarm time. If calendars supported the notion of location (besides simply providing a field to type it in), alarms could be automatically set based on how long it would take the user to reach the event.

  \item \textbf{Alarms on multiple devices.} When an alarm is set on multiple devices, each will go off at the exact same time without any knowledge of all the others. There is need to establish communication among the devices to present a single alarm to the user on the mutually-determined dominant device at the time.
\end{itemize}

\subsection{Supporting a Rich Variety of Event Types}

Users reported that not all events were equal: public events were merely for awareness, recurring events indicated that time was blocked out, and holidays were added to prevent accidental scheduling. From the users' point of view, each has different connotations, different visibility (public events should ideally fade out of sight when not required), and different types, number and intervals of alarms.

\begin{itemize}
  \item \textbf{Event templates.} A calendar system that supports event types can provide ways and means for users to create event templates and categories with different default settings al\-ong each of the dimensions outlined above. By having event templates, quick capture is supported as well. When much of the extra information about an event is pre-filled, data entry can be minimized to just the title of the event. Certain types of events have special metadata fields associated with them, e.g. conference call events contain the dial code, flight events contain airline and arrival/departure info. This could be easily achieved by event templates.

  \item \textbf{Showing/hiding public events.} While a few users said they added public events for informational purposes, others did not want public events (that they would not necessarily attend) to clutter their calendar. If calendars supported making certain event types visible or invisible on demand, the needs of both user groups could be met. Ag\-ain, by providing an option to keep all events in the same calendar, such a system would contribute to reducing information fragmentation.

\end{itemize}

\subsection{Reporting and Archival Support}

Report generation is a significant use of electronic calendars. Calendar software should have a way to generate reports and export information so that particular groups of events can be summarized in terms of when the meetings/events occurred, how many hours were devoted to them, and capture any notes entered in the calendar. One participant reported that he uses the search functionality in his calendar to obtain a listing of events related to a theme. This is used to get an idea of the number of hours devoted to particular activities and help to prepare an annual activity report.

\section{Discussion \& Future Work}

The paradox of encoding and remembering, as described in \cite{payne_1993_understanding}, was clearly evident in our data. Participants seem to over-rely on calendar artifacts to remember appointments, as seen in the setting of multiple alarms, printing of calendars for meetings, carrying a PDA everywhere, and calling their secretary to confirm events. The unfortunate side effect of sharing the management of a calendar with other people is that the primary user no longer goes through the personal encoding episode of entering the information. Some participants relied on administrative assistants to enter events in their calendars. At home, many participants relied on their spouses to maintain the calendar. Some participants even suggested the need to have an alarm for when events were added to their calendars. All of this points to a diminished opportunity for encoding the information that is entered into one's calendar. This makes it very difficult for participants to remember what is in their calendar, given that at times the scheduled events have never been seen before they occur.

On the other hand, the opportunity for rehearsal is greater today, if users take advantage of existing information dissemination and syndication techniques. For example, keeping a calendar on a desktop computer and publishing to an online calendar service such Google Calendar or Apple Mobile Me makes the calendar available in many other locations. Users can view their calendar on the web from any web browser, from mobile phones, or in the background on a desktop computer as part of widgets (tiny applications) such as Apple's Dashboard or Google Gadgets, or access it over a regular phone call \cite{perez-quinones_2004_youve}. So, the possibility of opportunistic rehearsal is afforded by current systems. We did not observe this in our data, as many of our users did not use these services. However, the paradox of encoding, rehearsal, and recall seems to be in need of future work so we can understand the impact of electronic calendar systems on human memory.

\section{Appendix}

\subsection{Survey Questions}
\subsubsection{Demographics}
\begin{itemize}
  \item Gender
  \item Occupation (Undergraduate Student, Graduate Student, Faculty, Staff, Industry)
  \item What is your age group?
\end{itemize}

\subsubsection{Calendar Use Basics}
\begin{itemize}
  \item Which devices do you own or use frequently?
  \item What computing-enabled calendars do you use?
  \item Do you use your computer to keep your calendar? If so, which program do you use for your main calendar management task on your desktop/laptop computer?
  \item If you own and/or use a PDA, which calendar program do you use on the PDA?
  \item Do you use an online calendar?
  \item What events do you record on your calendar?
  \item How often do you visit your calendar?
  \item How far ahead do you regularly look when you view your calendar?
  \item What would you consider your preferred view?
  \item If your calendar software includes a To-Do function, do you use it?
  \item Does your calendar software have a way to classify calendar events by categories? If so, how do you use this feature?
  \item Who changes and updates your calendar?
  \item How often do you add new events?
  \item Do you keep `proxies' (for example, post-its) or other notes that need to be entered in the calendar at a later time? 
  \item How long does it take for the proxy to make it into your main calendar?
\end{itemize}

\subsubsection{New Events}  

\begin{itemize}
  \item How frequently do you get events by phone (someone calls you) that go into your calendar?
  \item How frequently do you get events by e-mail (someone sends you email) that go into your calendar?
  \item How frequently do you get events in person (someone tells you of a meeting) that go into your calendar?
  \item By what other methods do new events arrive?
\end{itemize}

\subsubsection{More Calendar Use} 
\begin{itemize}
  \item What kinds of calendar failure (e.g. missed an event due to not entering it in calendar, scheduled conflicting events) have you experienced?
  \item What do you store in the calendar entry (check all that apply)?
  \item What other information do you record in your calendar?
  \item Do you use the alarms options of your calendar software?
  \item If you do use them, how much advance warning do you set your alarms for?
  \item If you do not use them, why not?
  \item If you are setting an alarm for an event, how many alarm notification do you set?
  \item What medium of notification do you use for your alarms?
  \item Please explain the role of an alarm?
\end{itemize}

\subsubsection{Use of PDAs}
\begin{itemize}
  \item Do you synchronize your PDA calendar with a desktop calendar?
  \item If so, how often?
  \item Which do you consider the `master' or main calendar?
  \item Do you make changes only on the desktop, thus keeping your PDA as a mobile/view-only aid?
  \item When you are at your desk, which device do you use to look up events?
  \item When you are at your desk, which device do you use to enter new events or change existing ones?
\end{itemize}

\subsubsection{Use of Paper Calendar Products}
\begin{itemize}
  \item Do you print your calendar?
  \item Which view do you print most frequently?
  \item How often do you print any of your views?
  \item Do you make changes (or write notes) in the printed copy?
  \item How often do you copy changes back to your computer?
\end{itemize}

\subsubsection{Use of a Wall Calendar}

\begin{itemize}
  \item Do you have a wall calendar in your home or office?
  \item If so, do you mark events there?
  \item If so, what types of events do you put on your wall-calendar?
  \item Do you copy these events to your `main' computerized calendar?
\end{itemize}

\subsubsection{Public Display of Calendar Information}
\begin{itemize}
  \item Do you keep a publicly available calendar (like posted office hours on your door, or in the department web page)?
  \item If so, how often do you update it?
  \item If you keep an online calendar, what privacy settings do you use?
\end{itemize}

\subsubsection{Sharing of Calendar}

\begin{itemize}
  \item Do you share your calendar with others?
  \item If so, with whom?
  \item Do you coordinate calendar events with your spouse, roommate, family?
\end{itemize}

\subsubsection{Phone Access to Calendar}
\begin{itemize}
  \item Do you use a service that provides access to your calendar over the phone?
  \item Would you like to use such a service?
\end{itemize}

\subsection{Interview Questions}

\begin{itemize}
  \item How important is your calendar to your work/daily activities?
  \item If you use multiple calendar systems, how often do you sync them? On a regular basis i.e.: every morning when you arrive at work, or as events occur?
  \item If you use multiple calendar systems, do you go back and forth between them during the day? Which do you use when and why?
  \item Do you use a to-do list?
  \item What events are on the to-do list versus scheduled as an event on the calendar? I.e.: presentation for a meeting are a to-do or is the meeting just scheduled on the calendar with a note in the entry that a presentation is due at that time?
  \item Do you keep ``proxies'' (sticky notes) or other notes that need to be entered in the calendar at a later time?
  \item What situations arise that make it necessary to use these proxies?
  \item Do you record public events on your calendar? (e.g. beginning of the Olympic games, holidays, election days, football dates, etc.)?
  \item What is your reasoning behind recording these public events?
  \item If you use alarms, what is the role of the warning? (i.e.: warning for a timed event or an appointment (such as History, 1-2pm Friday), warning to prepare or plan for an event, appointment, or deadline (such as History Test Friday), period events (Olympic games in session, football weekend))?
  \item Have you experienced any calendar failures? What kinds of calendar failures have you experienced? Please elaborate.
  \item Do you print your calendar? If so, do you change the printed copy?
  \item What are your reasons for making changes in the printed copy?
  \item Do you copy these changes back into your electronic calendar at any point?
  \item How long do you use this printed copy before printing out an updated one?
  \item Do you keep more than one calendar? (A wall calendar or other paper, computer, etc.)
  \item What criteria separates what goes on one or the other?
  \item Is there any overlap? Is one just a pared-down version of the other one or do they contain completely separate events?
  \item Do you coordinate calendar events with your spouse, roommate, family?
  \item If so, how do you go about doing that?
  \item Please explain any additional ways in which you use your calendar system.
  \item What are you habits as far as when you look at your calendar, how often, how far ahead do you look, how in-depth you examine events when you look, etc.
  \item Do you use a method of organization on a paper calendar that you cannot apply to an electronic calendar? (i.e.: specific types of events go into a specific area of the date box, highlighted events, etc)
  \item Is there anything else about your personal information management we have not covered?
\end{itemize}

\section{Acknowledgments}

The authors would like to thank all our survey participants and interviewees for their participation. We would also like to acknowledge Kyunghui Oh for assistance with preliminary analysis of the data collected, and Leysia Palen and Pardha Pyla for their comments on earlier drafts of this paper.

\end{document}